\documentclass[a4paper,11pt]{article}
\usepackage{jcappub} % for details on the use of the package, please see the JINST-author-manual
\usepackage{lineno}
\newcommand{\oiii}{[O{\sc\,iii}]}

\title{\boldmath Predicting Interloper Fraction with Graph Neural Networks}

\author[a,b]{Elena Massara,}
\author[c]{Francisco Villaescusa-Navarro,}
\author[a,b,d]{Will J. Percival}
\affiliation[a]{Waterloo Centre for Astrophysics, University of Waterloo, 200 University Ave W, Waterloo, ON N2L 3G1, Canada}
\affiliation[b]{Department of Physics and Astronomy, University of Waterloo, Waterloo, ON N2L 3G1, Canada}
\affiliation[c]{Center for Computational Astrophysics, Flatiron Institute, 162 5th Avenue, 10010, New York, NY, USA}
\affiliation[d]{Perimeter Institute for Theoretical Physics, 31 Caroline St North, Waterloo, ON N2L 2Y5, Canada}

\emailAdd{elena.massara.cosmo@gmail.com}

\abstract{Upcoming emission-line spectroscopic surveys, such as Euclid and the Roman Space Telescope, will be affected by systematic effects due to the presence of interlopers: galaxies whose redshift and distance from us are miscalculated due to line confusion in their emission spectra. Particularly pernicious are interlopers involving the confusion between two lines with close emitted wavelengths, like H$\beta$ emitters confused as \oiii, since those are strongly spatially correlated with the target galaxies. They introduce a particular pattern in the 3D distribution of the observed galaxy catalog that can shift the position of the BAO peak in the galaxy correlation function and bias any cosmological analysis performed with that sample. Here we present a novel method to predict the fraction of interlopers in a galaxy catalog, using Graph Neural Networks (GNNs) to learn the posterior distribution of the interloper fraction while marginalizing over cosmology and galaxy bias. The method is developed using simulations with halos acting as a proxy for galaxies. The GNN can infer the mean and standard deviation of the posterior distribution of interloper fraction using small-scale information that is usually not considered in cosmological analyses. The injection of large-scale information into the graph as a global attribute improves the performance of the GNN when marginalizing over cosmology.}

\begin{document}
\maketitle
\flushbottom

\section{Introduction}
\label{sec:intro}

Upcoming galaxy redshift surveys like Euclid\footnote{\url{https://www.euclid-ec.org}} and Roman\footnote{\url{https://roman.gsfc.nasa.gov/}} will observe emission-line galaxies up to high-redshift. They will use slitless spectroscopy to obtain galaxy spectra and identify one or more emission lines to measure the redshift of each galaxy. This technology will allow us to map the 3D distribution of galaxies with unprecedented detail, but the noisy spectra will require lots of attention to avoid systematics. One major issue concerns the presence of interlopers: objects for which the redshift has been wrongly estimated and, therefore, the distance has been miscalculated. This phenomenon can happen when lines in the galaxy spectrum are misidentified, which is more likely when spectra are noisy and when the redshift is estimated using a single emission-line. 

The main target of the Euclid spectroscopic survey is H$\alpha$ emission-line galaxies at redshifts $z=0.9 - 1.8$, whereas the High Latitude Spectroscopic Survey in Roman will observe H$\alpha$ emitters at redshifts $z\in[1,2]$, and \oiii\ emitters at $z\in[2,3]$. In Euclid, only the H$\alpha$ line will be used to calculate the galaxy redshifts, making the procedure prone to line confusion, which happens when the H$\alpha$ line is not visible and another strong line can be confused with the target. In Roman, redshifts are expected to be measured using at least two lines in each spectrum, a requirement that will discard many spectra where only one line is above the detection limit and will reduce the number of objects in the final catalog. The use of a single-line redshift identification would allow a much higher number density of galaxies and probe the large-scale structure in greater detail; however, such a procedure is more prone to line confusion and the presence of interlopers. 

In this paper, we consider the possibility of creating such a large catalog with the Roman Space Telescope, using single-line identification at high redshifts, where \oiii\ is the main target. Many lines can be confused as the target, and they can be classified into two main groups depending on the vicinity of the emitted wavelength of the detected line to that of the \oiii\ line. If they are far from each other in the spectrum, the interloper's inferred location is very far away from its true position, which is likely outside the redshift range of the main targets. As a consequence, this type of interlopers will only be very weakly correlated with the target population. On the other hand, if the two emitted wavelengths are close to each other in the spectrum, then the difference between the inferred and true distance to the interloper can be small, and the interloper population can be correlated to the target sample. An example of this second case is an H$\beta$ line confused as an \oiii\ line, resulting in the interloper distance being miscalculated by about $100~h^{-1}$Mpc. 

Both types of interlopers will modify the clustering properties of the full (target+interloper) sample and thus its two-point correlation function. Interlopers that are uncorrelated with the target sample have been studied in many papers (e.g. \cite{Pullen_2015,Gebhardt_2018,Addison_2018,Gong_2021}) and their effect on the correlation function can be easily modeled. On the other hand, interlopers that correlate with the target sample are more pernicious and more difficult to model. \cite{Massara_2021} studied the impact of \oiii-H$\beta$ interlopers on the correlation function, and in particular on the baryon acoustic oscillation (BAO) peak in it. They pointed out that the presence of this type of interlopers shifts the position of the BAO peak, introducing a systematic error in the BAO analysis. \cite{Setareh_2022} proposed a BAO fitting function that takes into account the presence of \oiii-H$\beta$ interlopers in a catalog (hereafter called BAO+$f_i$ fit), enabling an unbiased cosmological BAO analysis and a prediction for the fraction of objects in the catalog that are interlopers. Even if very successful, that model assumes that the target and interloper populations have the same galaxy bias, which is not guaranteed to be true for \oiii\ emitters and the galaxy population where H$\beta$ is the only detectable line; thus further improvements will be needed to accurately describe the impact of these interlopers in a BAO analysis. While these previous studies focused on the BAO analysis, we should remember that \oiii-H$\beta$ interlopers will affect other statistics of the large-scale structure beyond the two-point correlation function. 

In this paper, we develop a new method to predict the fraction of interlopers in a given catalog that are H$\beta$ emitters and confused as \oiii. Our method does not rely on the measurement and modeling of a summary statistic of the galaxy distribution, such as the two-point correlation function. Instead, it considers the full 3D distribution of galaxies by describing their positions with a mathematical graph and uses graph neural networks (GNNs)~\cite{Battaglia_2018} to extract the information about the interloper fraction. GNNs are designed to deal with sparse and irregular data, and have been applied in many areas of astrophysics (e.g.~\cite{Cranmer_2020,Villanueva-Domingo_2022,Villanueva-Domingo_2021,Villanueva-Domingo_2022b,DeSanti_2023,Shao_2023}). Since graphs encode the 3D spatial information beyond the two-point function, they allow us to use additional information to the two-point function to predict the fraction of interlopers in a catalog, and especially the information on small scales. We will show that thanks to these features GNNs can efficiently detect a subset of objects whose spatial clustering properties are different from the main sample. Moreover, while we present a GNN application to detect \oiii-H$\beta$ interlopers that will be a specific issue in Roman, GNNs can be used to recognize also other types of interlopers, such as those that are not correlated to the main sample and that will contaminate the spectroscopic catalog in Euclid. More generally, GNNs are promising tools to detect any subsets of objects exhibiting a spatial pattern different from a main sample. 

The paper is organized as follows: in Section~\ref{sec:interlopers} we discuss the spatial distortion introduced by \oiii-H$\beta$ interlopers and in Section~\ref{sec:GNN} we describe the data set used, how graphs are built, and the architecture and training procedure of our models. We present our results in Section~\ref{sec:results} and discuss the GNN approach used in this work in Section~\ref{sec:discussion}, before presenting the conclusions in Section~\ref{sec:conclusions}.

\section{Interlopers}
\label{sec:interlopers}

\begin{figure}[t]
\centering
\includegraphics[width=.99\textwidth]{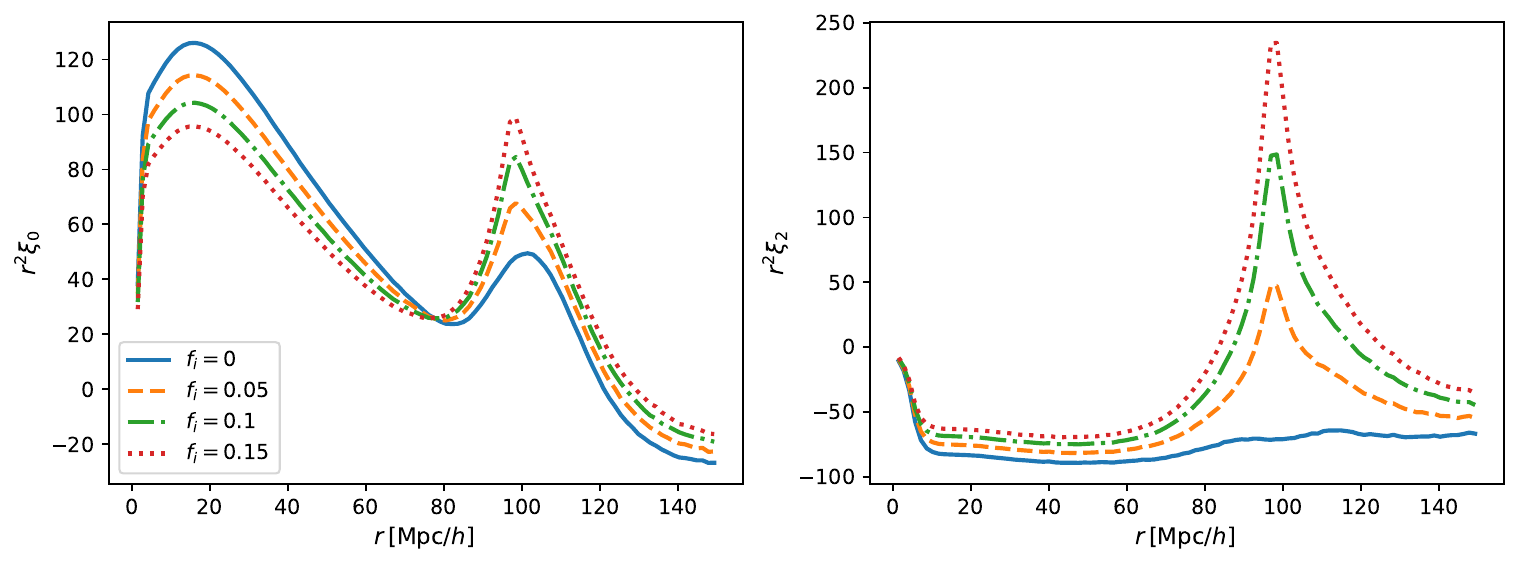}
\caption{Monopole (left) and quadrupole (right) of the two-point correlation function of halos measured from 1,000 Quijote simulations at redshift $z=1$ by~\cite{Setareh_2022}. Different colors and line styles show catalogs with different interloper fractions $f_i$.
\label{fig:xi}}
\end{figure}

An interloper is a galaxy located at the true redshift $z$ but considered to be at the wrong redshift $z'$. This mistake can happen when a line in the emission spectrum---where it appears at the observed wavelength $\lambda_{\rm o}$---is confused as another line having emitted wavelength $\lambda_{\rm e}^{\rm false}$ different from the true emitted wavelength $\lambda_{\rm e}^{\rm true}$. Subsequently, the redshift is converted into distance assuming a fiducial cosmology, thus a redshift misidentification leads us to infer a wrong distance for the interloper that is incorrect by
\begin{equation}
    \Delta d = d^{\rm false} - d^{\rm true} \simeq 
    \frac{c \, (1+z)}{H(z)}\left[ 1-\frac{\lambda_{\rm e}^{\rm true}}{\lambda_{\rm e}^{\rm false}}\right]\,,
\end{equation}
where $c$ is the speed of light and $H(z)$ is the Hubble parameter in the fiducial cosmology assumed to analyze the galaxy survey, i.e., to convert angles and redshifts to distances. 

When considering H$\beta$ emitters ($\lambda_{\rm e}^{{\rm H}\beta}=\lambda_{\rm e}^{\rm true}=486.1$nm) that are confused to be \oiii\ lines ($\lambda_{\rm e}^{\rm [O\,III]}=\lambda_{\rm e}^{\rm false}=500.7$nm), the offset between the true and wrongly inferred interloper position is equal to
\begin{equation}
\label{eq:displacement}
    \Delta d \simeq 87.41 \, \frac{1+z}{\sqrt{\Omega_\Lambda+\Omega_{\rm m}(1+z)^3}}\, h^{-1}{\rm Mpc}\,,
\end{equation}
for a $\Lambda$CDM model, where $\Omega_\Lambda$ and $\Omega_{\rm m}$ are the energy density of cosmological constant and matter in the assumed fiducial cosmology, respectively. This displacement does \emph{not} carry any cosmological information since it only depends on the fiducial cosmology, and it is equal to about $97~h^{-1}{\rm Mpc}$ in a Planck-like \citep{Planck2018} cosmology at $z=1$, the redshift that we consider in this paper\footnote{This choice is a balance between considering a high redshift where the \oiii-H$\beta$ interlopers will appear and the needing high number density of halos to perform the study}.
The offset is small enough that the true positions of the interlopers are in the same volume of the Universe covered by the target sample and the two populations are strongly correlated. When computing the two-point correlation function of the full sample, the correlation between the target galaxies and the interlopers appears as a peak centered at the scale of the displacement $\Delta d$, which is close to the BAO peak position at redshifts $z\sim 1-3$. The redshift and angular distances of the BAO peak depend on the true cosmology, and they are converted to comoving distances based on the fiducial cosmology. But in the BAO case, unlike the interloper offset, they still conveys cosmological information from redshift and angular distances. The match between the BAO position and the interloper offset therefore strictly relies on the choice of the fiducial cosmology, the emission lines considered for the interloper and target galaxies, and the redshift of the target sample. 

Figure~\ref{fig:xi} shows measurements of the monopole (left) and quadrupole (right) of the two-point correlation function in halo catalogs created with different fractions of interlopers $f_i$ at $z=1$ (see Section~\ref{subsec:data} for details on how these catalogs are generated). 
As the interloper fraction increases, the BAO peak in the monopole appears to be more and more enhanced, broadened, and shifted towards smaller separations $r$. Moreover, the quadrupole exhibits a peak at around $r=97~h^{1}$Mpc, whose height increases with the interloper fraction, showing that interlopers create an anisotropic pattern in the 3D distribution of the halos. Since the BAO position is used as a standard ruler in cosmology, any source that shifts its position will introduce a systematic bias in the cosmological analysis: the larger the interloper fraction, the larger the shift of the BAO position, and the larger the systematic introduced in the analysis~\cite{Massara_2021}. \cite{Setareh_2022} developed the BAO+$f_i$ fit, a BAO fitting function that takes into account and models the effect of interlopers, however the 3D distortion introduced by this type of interlopers is going to affect other statistics measured from a contaminated catalog. In this paper, we develop a method to predict the \oiii-H$\beta$ interloper fraction directly at the field level, and this methodology can be in principle used to detect the fraction of different types of interlopers. We do this by focusing on using the small-scale information in order to separate the measurement of interlopers from the cosmological measurements. Because of the shift in position, the measured small-scale clustering around the interlopers will be very different from that of the unshifted galaxies, but is difficult to model. It is this information that we hope the GNN will be able to extract.

\section{Graph Neural Network}
\label{sec:GNN}
\subsection{Data sets}
\label{subsec:data}
In this work, we use dark matter halos as a proxy for galaxies to test our method. Both galaxies and halos are biased tracers of the underlying matter field, with bias schemes that can be tuned by the choices made to build their population. To test the feasibility of our GNN method, we do not build galaxy catalogs with particular emission line distributions that have realistic fractions of interlopers in them. We instead generate a sufficiently large data set with enough variation in the objects' bias, the underlying cosmology, and the fraction of interlopers, so that the training set can be used to train a flexible enough model that can predict the fraction of interlopers effectively marginalizing over the other unknowns. This is sufficient to test the main idea of the paper, that Graph Neural Networks can help us predict the number of interlopers in a catalog. However, a sufficient level of detail in the creation of these catalogs will be necessary to deploy the GNN method to the analysis of real data. In particular, galaxy catalogs will need to be used since they exhibit the Fingers-of-God effect on small scales~\cite{FoG}. Moreover, the effect of angular selection, completeness, and other observation systematics will need to be assessed and eventually introduced at the training stage level. 

We use halo catalogs from the Quijote simulations \cite{quijote}, a suite of thousands of N-body simulations built to perform Fisher analysis and train deep-learning models. The products we use were obtained in the following way. The initial conditions of the simulations were generated at redshift $z=127$ using second-order Lagrangian perturbation theory, and dark matter particles were evolved through the present time using the TreePM code Gadget-III; snapshots at redshifts $z=0, 0.5, 1, 2, 3$ were saved and used to identify halos via the Friends-of-Friends (FoF) algorithm~\citep{Davis_1985} with linking length parameter $b = 0.2$: All halos with more than 20 particles were saved into halo catalogs. All the simulation boxes have a cubic volume equal to $1~h^{-3}{\rm Gpc}^{3}$ and contain $512^3$ cold dark matter particles. In our analysis we made use of two different sets of simulations:

\begin{itemize}
\item {\tt SET1}: 100 simulations at the so-called fiducial cosmology, a flat $\Lambda$CDM cosmology with matter density parameter $\Omega_{\rm m}=0.3175$, baryon density parameter $\Omega_{\rm b}=0.049$, dimensionless Hubble constant $h=0.6711$, spectral index $n_s=0.9624$, linear matter fluctuation amplitude $\sigma_8=0.834$, and sum of neutrino masses $M_\nu=0$ eV. With this cosmology, the minimum halo mass present in the catalogs is equal to $1.31\times10^{13}~h^{-1}$M$_\odot$. 
\item {\tt SET2:} 155 simulations selected among 2,000 simulations whose cosmological parameters are arranged in a Latin Hypercube (LH) configuration. The selected boxes cover a fraction of the full LH volume around its center and have parameters $\Omega_{\rm m}\in\left[0.18-0.42\right]$, $\Omega_{\rm b}\in\left[ 0.038-0.062\right]$, $h\in\left[ 0.58-0.82\right]$, $n_s\in\left[ 0.88-1.12\right]$, $\sigma_8\in\left[ 0.68-0.92\right]$. 
\end{itemize}

We consider the halo catalogs at redshift $z=1$ for all of these simulations and use the ones in {\tt SET1} to build a model that predicts the fraction of interlopers at fixed cosmology and fixed/varying halo bias, while we employ the catalogs in {\tt SET2} to build a more flexible model that can marginalize over both cosmology and halo bias, having them both varying in the training step. We analyze both sets assuming a fiducial cosmology, corresponding to the cosmology used to run {\tt SET1}. Because of this, the boxes in {\tt SET2} are stretched to account for the difference between the cosmology of that simulation and the assumed one. We consider the line-of-sight to be along the $\hat{z}$ direction so that the 3D halo comoving positions $(x_x,x_y,x_z)$ are mapped into the stretched coordinates $(x'_x,x'_y,x'_z)$ via
\begin{equation}
    \label{eq:stretch1}
    x'_x = \frac{x_x}{\alpha_\perp}, \qquad x'_y = \frac{x_y}{\alpha_\perp}, \qquad x'_z= \frac{x_z}{\alpha_\parallel}
\end{equation}
using the parameters 
\begin{equation}
    \label{eq:stretch2}
    \alpha_\parallel = H(z) / H_{\rm sim}(z), \qquad \alpha_\perp = D_{A,{\rm sim}}(z) / D_A(z) 
\end{equation}
where the subscript ``sim" denotes the parameter in the cosmology used to run the simulation and absence of subscript denotes parameters in the assumed fiducial cosmology, and $D_A$ is the angular diameter distance.

Due to constraints from the memory of the GPUs, we cannot build a graph using the full halo catalog in a simulation box, thus we crop them into sub-boxes of size that varies depending on the task. Interlopers are introduced in each sub-box by randomly selecting a fraction $f_i$ of halos and shifting them along the line-of-sight $\hat{z}$ by the distance predicted in Equation \ref{eq:displacement}. The fraction of interlopers $f_i$ assigned to each sub-box is a number randomly sampled within the interval $f_i=\left[ 0.0-0.2\right]$.

\subsection{Graphs}

We describe the 3D distribution of the halos using graphs, where halos are represented by nodes that can be connected to each other via edges depending on their separation: only nodes closer than the linking radius $r_{\rm link}$ are connected via edges. The value of the linking radius is a hyperparameter of our GNN model and will be chosen to maximize the performance of the model. 

Nodes, edges, and the graph itself can be labeled with attributes carrying information that describes them. For the purpose of this study, each graph will encode knowledge about the halo spatial distribution only; any information about halo properties, such as their mass, concentration, or history, is not used. Moreover, it is important to create deep-learning models that respect the symmetries of the problem they are going to describe \cite{Villar_2023}. The observed Universe exhibits a cylindrical symmetry\footnote{The Universe is statistically isotropic (invariant under rotations), however we observe redshifts rather than distances. The galaxy redshift is not only determined by the Hubble flow, hence its distance to us, but also by the peculiar velocity of the galaxy along the line-of-sight $\hat{z}$. This causes a distortion when converting redshifts to distances known as redshift space distortion, which introduces a special direction and breaks the isotropy. Moreover, when converting redshift and angles into distances, a fiducial cosmology needs to be assumed. If that is different from the cosmology of the Universe (or the simulation box considered), additional anisotropic distortions are introduced in the dataset.} and we construct graphs that respect it by assigning the spatial information to the edges instead of the nodes: each edge attribute is composed by three scalars that describes the 3D vector connecting two nodes via
\begin{equation}
\label{eq:x}
{\bf e}_{ij} = \left[ \, r_{\parallel} = \frac{{\bf d}_{ij}\cdot \hat{z}}{r_{\rm link}} \,,
\qquad
r_{\perp} = \frac{ | {\bf d}_{ij} \times \hat{z}|}{r_{\rm link}} \,,
\qquad
\cos\theta=\frac{{\bf v}_{i\perp} \cdot {\bf v}_{j\perp}}{|{\bf v}_{i\perp}||{\bf v}_{j\perp}|} \, \right],
\end{equation}
where ${\bf d}_{ij} = {\bf v}_i-{\bf v}_j$ is the vector connecting the two nodes $i$ and $j$ at the beginning and end of the edge ${\bf e}_{ij}$, and ${\bf v}_\perp={\bf v}\times \hat{z}$ is the component of ${\bf v}$ perpendicular to the line of sight. 

Initially, an entire simulation box is cropped into smaller subvolumes, along the $\hat{x}$ and $\hat{y}$ directions (not cut in the $\hat{z}$ direction to better capture interloper effects), to have a representative sample of halos but small enough to fit into the GPU memory. We build a graph from the halos in each subvolume using the above procedure. 

In some cases, we consider a global attribute to describe the full graph. Depending on the cases, it will be given by the five cosmological parameters or the monopole of the halo power spectrum measured in the sub-box.

\subsection{GNN architecture}

We use the graphs built from the halo catalogs to train a GNN that predicts the fraction of interlopers in the catalog. The GNN consists of multiple blocks that update the node, edge, and global attributes while maintaining the structure of the graph. These blocks are Metalayers~\cite{Battaglia_2018} that use a message-passing scheme to effectively flow the information from node to node and update attributes. At the end, a global pooling and a multi-layer perceptron (MLP) are added to compress the graph information to the desired size of the output: the prediction for the fraction of interlopers in a graph.  

Each GNN block takes as input a graph, updates its edge, node, and global attributes, and outputs the updated graph. Thus, even if the initial graphs do not have node attributes, the GNN blocks will assign and update them via message passing. Each block $l$ is composed of the following elements:
\begin{itemize}
    \item The edge model that updates each input edge attributes ${\bf e}_{ij}^{(l-1)}$ to the output ${\bf e}_{ij}^{(l)}$:
    \begin{equation}
    \label{eq:edge_model}
       {\bf e}_{ij}^{(l)}=\phi_{l}\left(\left[ {\bf n}_i^{(l-1)},{\bf n}_j^{(l-1)},{\bf e}_{ij}^{(l-1)}\right]\right) \,,
    \end{equation}
    \item The node model that updates the node attributes:
    \begin{equation}
    \label{eq:node_model}
       {\bf n}_{i}^{(l)}=\psi_{l}\left(\left[ {\bf n}_i^{(l-1)},\bigoplus_{j\in {\mathcal{N}}_i}{\bf e}_{ij}^{(l)},{\bf u}\right]\right) \,,
    \end{equation}
\end{itemize}
where $\phi$ and $\psi$ are MLPs, ${\bf u}$ is the global attribute (when specified), and $\bigoplus$ is a permutation invariant aggregation operator applied to the edges ${\bf e}_{ij}$ with  $j\in{\mathcal{N}}_i$ and ${\mathcal{N}}_i$ being all the nodes that are neighbours connected via edges to the node ${\bf n}_{i}$. We consider three different types of aggregations---summation, maximum value, mean---and concatenate all of them in equation \ref{eq:node_model}. In this way each node is updated depending on the values of the node attributes of its neighbours and the attribute of the edges connecting them, allowing for a flow of information. The number of GNN blocks is a hyperparameter chosen to maximize the performance of the model; it determines how many times the message passing is performed and the attributes are updated.

After the GNN blocks, the information contained in the graph ${\mathcal{G}}$ is compressed using one last aggregation operation $\bigoplus$ performed on all nodes. The result is concatenated to the global attribute, if present in the graph, and passed to a final MLP $\tau$ to output the vector
\begin{equation}
    \label{eq:pulling}
       {\bf y}=\tau\left(\left[ \bigoplus_{i\in {\mathcal{G}}}{\bf n}_{i},{\bf u}\right]\right) \,.
    \end{equation}
The MLP $\phi$, $\psi$, and $\tau$ are built using two fully connected layers with ReLu activation function. The number of GNN blocks and neurons per fully connected layer are hyperparameters to optimize. 

\subsection{Training procedure}
 
Our aim is to build a GNN model that can infer the fraction of interlopers $f_i$ in a catalog $\mathcal{G}$. Our model predicts the mean and standard deviation of the interloper posterior distribution without making any assumption about its shape. Thus, it outputs the vector ${\bf y}(\mathcal{G}) = [\mu(\mathcal{G}),\sigma(\mathcal{G})]$, with 
\begin{equation}
    \label{eq:mean}
\mu(\mathcal{G}) = \int df_i \, p(f_i|\mathcal{G}) f_i
\end{equation}
\begin{equation}
    \label{eq:std}
\sigma(\mathcal{G}) = \left[ \int df_i \, p(f_i|\mathcal{G}) (f_i-\mu)^2 \right]^{1/2}
\end{equation}
being the mean and standard deviation of the marginalized posterior distribution $p(f_i|\mathcal{G})$,
\begin{equation}
    \label{eq:posterior}
p(f_i|\mathcal{G}) = \int d\theta_1...d\theta_n \, p(f_i,\theta_1,...,\theta_n|\mathcal{G})\,,
\end{equation}
where $\theta_1,...,\theta_n$ are cosmological parameters or parameters describing the galaxy/halo bias scheme. Here $\sigma(\mathcal{G})$ represents the aleatoric error alone since we do not include the small epistemic error (see Section~\ref{sec:epistemic} for a quantitative analysis). 

In order to train such a model, we implement the loss function \cite{Jeffrey_2020,Camels_2022}
\begin{equation}
    \label{eq:loss}
       \mathcal{L}=\log \left\{ \sum_{j\in {\rm batch}} (f_{i,j}-\mu_j)^2\right\}+
       \log \left\{ \sum_{j\in {\rm batch}} 
        \left[(f_{i,j}-\mu_j)^2-\sigma_j^2\right]^2\right\}.
    \end{equation} 
whose minimization is equivalent to solving for the mean and standard deviation of the posterior distribution (see \cite{Jeffrey_2020, Villaescusa_2020}). We divide the data into training, validation, and test sets with a 80/10/10 split ratio. In the training stage, we minimize the loss function using the {\sc Adam} optimizer~\citep{kingma2017}, with values for the learning rate and weight decay that we treat as hyperparameters to be optimized. We train the models for at least 1,000 epochs and choose the model with the best validation loss. The optimization of the hyperparameters (learning rate, weigh decay, number of GNN blocks, number of neurons in $\phi$, $\psi$, and $\tau$, and linking radius $r_{\rm link}$) is performed using the {\sc Optuna} package~\cite{Akiba_2019} with at least 100 trials, each of those consisting in the training of a model with a specific choice for the value of hyperparameters. We select the GNN model with hyperparameters that give the best validation loss after training. Then, we determine the performance of the selected GNN model using the test set.

\subsection{Accuracy metrics}

We quantify the performance of our GNN models using different metrics, all applied to the test sets. We consider 
\begin{itemize}
    \item The root mean square error 
\begin{equation}
    \label{eq:RMSE}
{\rm RMSE}= \, \sqrt{< (\mu-f_i)^2>}
\end{equation}
with $<...>$ indicating the mean among the test set, which quantifies the precision of the model---the lower the $RMSE$ the more precise the model is.     
\item The coefficient of determination 
\begin{equation}
    \label{eq:R2}
{\rm R}^2 = 1-\frac{<(\mu-f_i)^2>}{<(f_i-<f_i>)^2>}
\end{equation}
that measures the accuracy of the model. It is limited to be $R^2\leq 1$: the closer it is to 1 the more accurate the model is. A value close to 0 indicates that the model is not properly trained and it can only predict the mean of the training set, while a negative value for $R^2$ denotes that the model is performing even worse than that.

\item An estimation for the bias
\begin{equation}
    \label{eq:bias}
b=<\mu-f_i>.
\end{equation}

\item An estimation for how well the standard deviation of the posterior distributions is determined by the GNN model,
\begin{equation}
    \label{eq:chi2}
\chi^2=\,\left<\left[\frac{\mu-f_i}{\sigma}\right]^2\right>\,.
\end{equation} 
A $\chi^2$ close to 1 indicates that the standard deviations are properly predicted.

\end{itemize} 
\section{Results}
\label{sec:results}
\subsection{Fixed cosmology and halo bias}
\label{sec:inference_fid_fixb}
We first consider the simplest case where all catalogs have the same cosmology, the fiducial cosmology of the Quijote simulations, and are built using {\tt SET1}. Moreover, we maintain all halos in the catalogs, so that the halo bias is also shared among the whole data set. This will allow us to build a GNN model at fixed cosmology and fixed halo bias scheme, where the posterior distribution of the fraction of interlopers $p(f_i|\mathcal{G})$ in equation ~\ref{eq:posterior} is not marginalized over any other parameter. 

In this case, we crop sub-boxes of size $150\times 150\times 1000\,(h^{-1}{\rm Mpc})^3$ along the $\hat{x}$, $\hat{y}$, and $\hat{z}$ directions, respectively, and with the line-of-sight assumed to be the $\hat{z}$ axis. Depending on the sub-box, a different fraction of halos are selected to represent interlopers and are displaced by $97\,h^{-1}$Mpc along the $\hat{z}$ direction, while taking into account the boundary conditions of the simulation box along the $\hat{z}$ axis. As a result, these sub-boxes contain about 4,500 objects. We consider both the catalogs in real (no velocity added) and redshift space, and perform two different studies on these two types of data sets to determine the best values for the hyperparameters. 

In real space, the trained model with the best validation loss has $N_{\rm block}=1$, $N_{\rm hid}=42$, $l_r=1.1\times 10^{-4}$, $w_d=4.9\times 10^{-3}$ and $r_{\rm  link} = 11.5\,h^{-1}$Mpc; in redshift space the equivalent best model has $N_{\rm block}=1$, $N_{\rm hid}=35$, $l_r=3.8\times 10^{-4}$, $w_d=10^{-2}$ and $r_{\rm  link} = 11.68\,h^{-1}$Mpc. The single GNN block and the low values for $r_{\rm  link}$ indicate that the GNN models are using small-scale clustering properties to determine the fraction of interlopers. As expected, there is lots of information on small scales, since the number of pairs and their spatial distribution change depending on the number of interlopers in the catalog and graph. 

\begin{figure}[t]
\centering
\includegraphics[width=.75\textwidth]{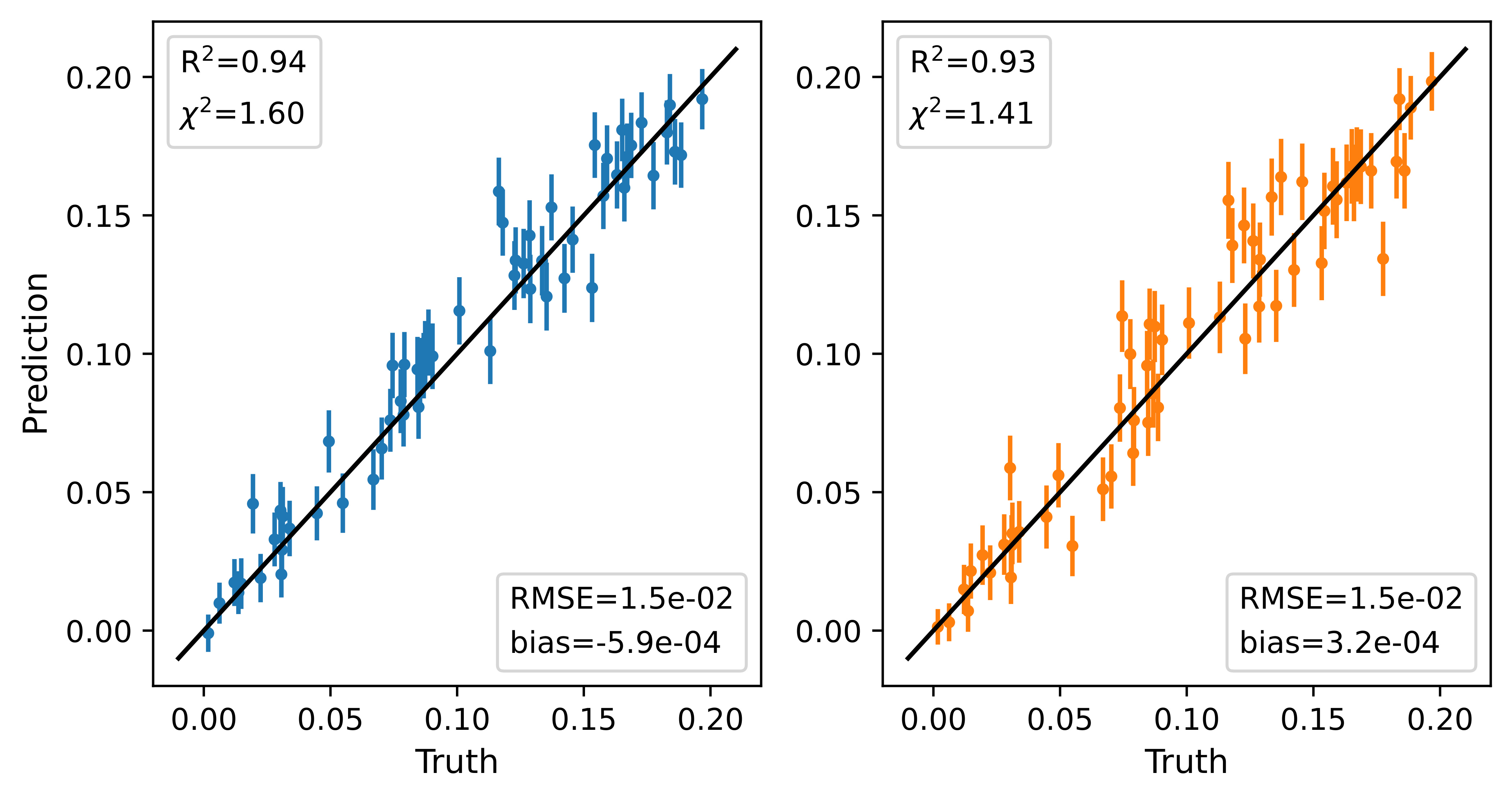}
\caption{Likelihood-free inference of the fraction of interlopers in halo catalogs sharing the same cosmology and same halo bias. Left panel shows the results in real space and right panel displays the results in redshift space.
\label{fig:inference_fid_fixb}}
\end{figure}

Figure~\ref{fig:inference_fid_fixb} shows the inference performed on the test set using the two models: real space (left) and redshift space (right). The $x$-axis indicates the true value $f_i$, while the $y$-axis denotes the prediction of the model. The black line shows the values $y=x$, indicating when the prediction matches exactly the true value. The blue points denote the predicted mean and the error bars denote the predicted standard deviation in each test catalog. By eyes, we can see that the error bars follow the black diagonal line, and no bias is present, as confirmed by the values of $b\sim0$ reported in the figure. Moreover, $\chi^2=1.6,1.4$ for the real and redshift space respectively, indicating that the standard deviation is slightly under-predicted by the two models, probably because we are not taking into account the epistemic error of the GNN (see~\ref{sec:epistemic} for more details). The models estimate the interloper fraction with a precision of $\pm0.015$, consisting of an error equal to $15\%$ on the mean value $f_i = 0.1$ of the considered interloper fraction range. This precision is obtained using a volume equal to $0.0225\,h^{-3}{\rm Gpc}^3$, which is a very small fraction of the \oiii\ survey in the Roman space telescope and a volume thousands of times smaller than the one considered in \cite{Setareh_2022}. 

We can compute the $RMSE$ for the analysis in \cite{Setareh_2022} using the results reported in their Figure 1 for the same displacement considered in this paper: $\Delta d=97~h^{-1}$Mpc. We obtained $RMSE=3.1\times10^{-3}$, which rescaled to the volume considered in this section ($0.0225~h^{-3}{\rm Mpc}^3$) is equal to $0.65$. Therefore, it is $40$ times larger than the $RMSE$ we obtain with the GNNs; in other words, the GNN models are 40 times more precise in predicting the interloper fraction than the BAO+$f_i$ fit. We can also compare the bias in the two methods: the one from the BAO+$f_i$ model fit in~\cite{Setareh_2022} is equal to $-2.1\times 10^{-3}$, which is an order of magnitude larger than the bias obtained with GNNs\footnote{We do not rescale the bias error with the volume}. However, we should note that the comparison is not truly fair since the GNN has been trained at fixed cosmology, whereas the BAO+$f_i$ fit includes the possibility for a variation in cosmology via the dilation parameters. 

\subsection{Fixed cosmology and varied halo bias}
\label{sec:inference_fid_varyb}

\begin{figure}[!t]
\centering
\includegraphics[width=.75\textwidth]{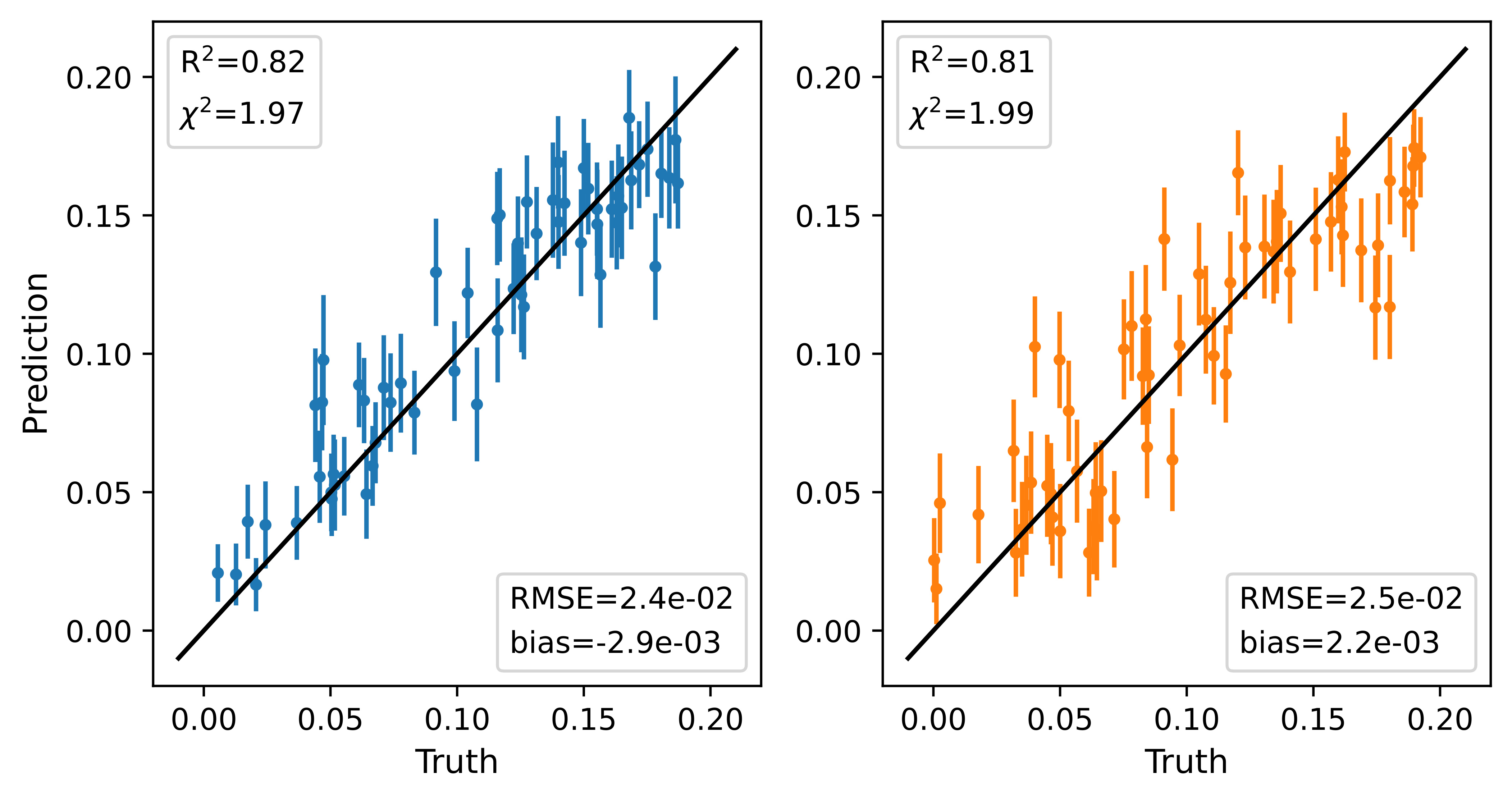}
\caption{Likelihood-free inference of the fraction of interlopers in redshift space halo catalogs sharing the same cosmology but having \emph{different halo bias}. 
Left panel: Each catalog is obtained from a volume equal to $150\times 150 \times 1,000 \, (h^{-1}{\rm Mpc})^3$ and only $N_{\rm halo}$ more massive halos are considered, where $N_{\rm halo}$ varies between 600 and 4,200.   
Right panel: the number of halos is fixed to $N_h=4,500$, with halos being randomly selected from those with mass larger than $M_{\rm h,min}$ in a volume equal to $250\times 250 \times 1,000 \, (h^{-1}{\rm Mpc})^3$. $M_{\rm h,min}$ randomly varies in the interval $1.31-1.97\times 10^{13}\, h^{-1}M_{\odot}$, to obtain different halo biases in each catalog, while maintaining the same number density.  
\label{fig:inference_fid_varyb}}
\end{figure}

Using the 100 realizations at the fiducial cosmology in {\tt SET1}, we can build catalogs with different halo biases. In this case, the posterior distribution $p(f_i|\mathcal{G})$ of which we aim to learn the mean and the standard deviation is marginalized over the halo bias scheme. We consider two different strategies to implement the bias scheme. The first method involves selecting the $N_{\rm halo}$ most massive halos in a catalog, where $N_{\rm halo}$ is a random number---we will refer to this method as \emph{varied-$N_{\rm halo}$}. Small values for $N_{\rm halo}$ correspond to the selection of only very massive halos and the creation of a halo catalog that has high bias. On the other hand, large values of $N_{\rm halo}$ involve the selection of low-mass halos as well and yield a population that is less biased. The second method concerns (1) randomly selecting the minimum halo mass $M_{\rm h,min}$ to include in the catalog and (2) randomly sub-sampling to a chosen number $N_{\rm halo}$ all the halos that are more massive or as massive as the chosen limit. In this way, we can have catalogs with different halo biases but sharing the same number density of objects---we will refer to this method as \emph{fixed-$N_{\rm halo}$}. 

The implementation of the first method (\emph{varied-$N_{\rm halo}$}) is done by cropping sub-boxes of size $150\times 150\times 1000\,(h^{-1}{\rm Mpc})^3$ along the $\hat{x}$, $\hat{y}$, and $\hat{z}$ directions respectively, and randomly selecting $N_{\rm halo}$ halos, with $N_{\rm halo}$ ranging between 6,000 and 4,200. This corresponds to a variation of the minimum halo mass within the range $1.31\times 10^{13}-5\times 10^{13}\,h^{-1}M_{\odot}$ and a variation in halo bias equal to a factor of 2. Interlopers are implemented by selecting a different fraction of halos in each catalog that are displayed by $97\, h^{-1}$Mpc along the line-of-sight. This procedure is repeated twice for each sub-box to augment the number of data. The model with the best validation loss trained using this data setup has hyperparameters $N_{\rm block}=2$, $N_{\rm hid}=66$, $l_r=9.4\times 10^{-5}$, $w_d=6.46\times 10^{-3}$ and $r_{\rm link} =17.96\, h^{-1}$Mpc, and its performance of the test set is shown in the left panel of Figure~\ref{fig:inference_fid_varyb}. 

To implement the second method (\emph{fixed-$N_{\rm halo}$}) we obtain sub-boxes of size $250\times 250\times 1000\,(h^{-1}{\rm Mpc})^3$ along the $\hat{x}$, $\hat{y}$, and $\hat{z}$ directions respectively. We choose to work with larger volumes because they allow us to have in each sub-box a number of halos larger than $N_{\rm halo}=4,500$, once only those with mass $M \leq M_{\rm h,min}$ are selected. We sample $M_{\rm h,min}$ within the range $1.31\times 10^{13}\leq M_{\rm h,min}\leq 1.97\times 10^{13}\,h^{-1}M_{\odot}$, which corresponds to a variation in halo bias at the level of $20\%$. Analogously to the \emph{varied-$N_{\rm halo}$} method, interlopers are then introduced in each sub-box by shifting a selected number of objects along the line-of-sight, and the procedure is repeated twice in each sub-box to augment the number of data. In this case, the model with best loss has $N_{\rm block}=1$, $N_{\rm hid}=41$, $l_r=4.9\times 10^{-4}$, $w_d=1.7\times 10^{-5}$ and $r_{\rm link} =25.05\, h^{-1}$Mpc. Results showing the model performance on the test set are displayed in the right panel of Figure~\ref{fig:inference_fid_varyb}

In both cases, the models are unbiased and are able to extract information about the fraction of interlopers from small scales. All metrics are very similar, with $\chi^2$ being close to 2, meaning that the standard deviations are under-predicted. When we compare these results with those in Section~\ref{sec:inference_fid_fixb}, we can see that allowing the halo bias to vary across different halo catalogs yields worse precision: the fraction of interlopers is determined with a precision of $\pm 0.025$ over the full range considered. The value of $R^2$ is also smaller than in the case of fixed halo bias. A comparison with the BAO+$f_i$ fit method in~\cite{Setareh_2022} shows that GNN models that marginalize over halo bias are about $15-25$ times more precise in predicting the interloper fraction than the BAO model, depending on the method used to introduce a variety of halo bias schemes in the training set (they consider different volumes), but they exhibit a similar level of bias in the prediction.

\subsection{Varied cosmology and halo bias}
\label{sec:inference_LH}

\begin{figure}[!h]
\centering
\includegraphics[width=.75\textwidth]{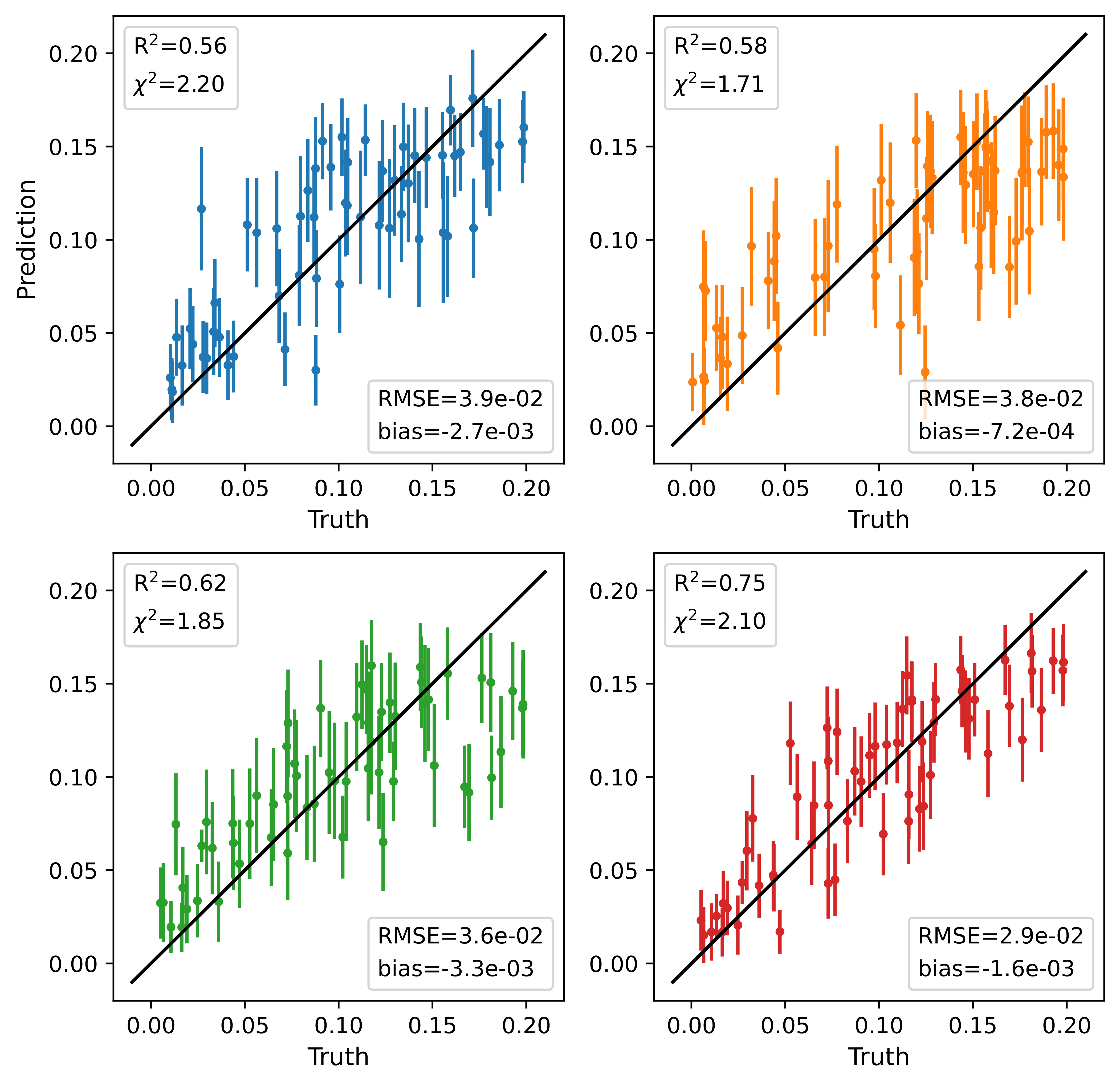}
\caption{Likelihood-free inference of the fraction of interlopers in redshift space halo catalogs with different cosmologies and halo bias. 
Top left panel: Each catalog is obtained from a volume equal to $150\times 150 \times 1,000 \, (h^{-1}{\rm Mpc})^3$ and only $N_h$ more massive halos are considered, where $N_h$ varies between 600 and 4,200. Top right panel: the number of halos is fixed to $N_h=4,500$, with halos being randomly selected from those with mass larger than $M_{min,h}$ in a volume equal to $250\times 250 \times 1,000 \, (h^{-1}{\rm Mpc})^3$. $M_{min,h}$ randomly varies in the interval $\left[1.31-1.97\right]\times 10^{13}\, h^{-1}M_{\odot}$, in order to obtain different halo biases in each catalog, while maintaining the same number density. Each simulation box in the selected area of the Latin Hypercube is used three times. Bottom left panel: the number of halos is fixed, and the halo mass cut and volume cut are the same as in the top right panel, and each simulation box in the selected area of the LH is used three times. However, here each graph has a global attribute that is the monopole of the measured power spectrum up to $k=0.3\,({\rm Mpc})^{-1}h$. 
Bottom right panel: the number of halos is fixed, and the halo mass cuts and volume cuts are the same as in the top right panel, but each simulation box in the LH is used three times. Moreover, the graphs have a global attribute with guess values for the cosmology randomly drawn from a normal distribution centered at the values of the true cosmology of the simulation and with variance equal to the 1$\sigma$ uncertainties from Planck. 
\label{fig:inference_LH}}
\end{figure} 

In order to apply a GNN to real data, the model needs to predict the mean and standard deviation of the poster distribution marginalized over halo bias and cosmology. We train such a model by building halo catalogs from {\tt SET2}, the LH of the Quijote suite. We implement the two ways described in the previous section to obtain differently biased populations starting from the same box or sub-box. The only difference with the previous section is that here each simulation box is at a different cosmology, thus each box has been stretched via Equations~\ref{eq:stretch1} and~\ref{eq:stretch2} and reshaped into a cube using boundary conditions before being cropped to the desired volume. The value for the interloper shift depends on the assumed fiducial cosmology via Equation~\ref{eq:displacement}, but it does not depend on the true cosmology of the simulations. Thus it does not carry any meaningful cosmological information and its value is the same in all the simulations of the LH. The performances of the best validation model on the test set are shown in Figure~\ref{fig:inference_LH} for the \emph{varied-$N_{\rm halo}$} method (top left) and \emph{fixed-$N_{\rm halo}$} method (top right). The corresponding hyperparameters are $N_{\rm block}=2$, $N_{\rm hid}=24$, $l_r=1.1\times 10^{-4}$, $w_d=1.1\times 10^{-4}$, $r_{\rm  link} =28.5\,h^{-1}{\rm Mpc}$ for the first method and $N_{\rm block}=2$, $N_{\rm hid}=39$, $l_r=1.2\times 10^{-4}$, $w_d=1.0\times 10^{-6}$, $r_{\rm  link} =29.9\,h^{-1}{\rm Mpc}$ for the second. The metrics on the performance of the two models are very similar and show a decrease in precision when allowing for the data set to have different cosmologies. Compared to previous cases at fixed cosmology, the accuracy went down with $R^2$ drastically decreasing to $0.56-0.58$, and the precision became worse and equal to $\pm 0.039$. However, these results show that GNNs can achieve a precision $10-17$ times better than the BAO+$f_i$ fit analysis in~\cite{Setareh_2022}, with both methods allowing for the cosmology and halo bias to vary.

The studies performed to understand the best values for the hyperparameters suggest the need to use larger linking radii, but that is not possible due to the limitation of the GPU memory on the system used. This suggests that there might be a degeneracy between cosmology, halo bias, and the fraction of interlopers on small scales. In order to add information coming from large scales, we consider adding a global attribute to the graphs. 

First, we implement as global attribute the monopole of the halo power spectrum measured from each halo catalog up to a maximum wavelength equal to $k_{\rm max}=0.3\,h\,{\rm Mpc}^{-1}$. We select different halo biases using the \emph{fixed-$N_{\rm halo}$} method because its setup allows us to use larger volumes than with the \emph{varied-$N_{\rm halo}$} scheme. However, even if the sub-boxes are larger, we do not include higher-order multipoles to the global attributes because their volume is still relatively small and higher-order multipoles are dominated by cosmic variance. The results using graphs with $N_{\rm halo}=4,500$ and a GNN model that used the monopole of the power spectrum as global attribute is shown in the bottom left panel of Figure~\ref{fig:inference_LH}. The best model has $N_{\rm block}=2$, $N_{\rm hid}=27$, $l_r=4.0\times 10^{-5}$, $w_d=1.3\times 10^{-4}$, $r_{\rm  link} =26.4\,h^{-1}{\rm Mpc}$. Adding this global attribute improves all metrics used to quantify the performance of the model, however, these improvements are relatively minor. 

Second, we consider as global attribute five numbers that are a guess for the values of the five cosmological parameters of the underlying cosmology. We guess them by randomly sampling the marginalized posterior distribution for these five parameters obtained by the Planck collaboration~\cite{Planck2018}, after shifting the posterior so that it is centered at the right value of a given parameter for the simulation considered. We train GNN models with this global attribute and chose the one with best validation loss, having $N_{\rm block}=2$, $N_{\rm hid}=27$, $l_r=1.3\times 10^{-4}$, $w_d=6.7\times 10^{-4}$, $r_{\rm  link} =22.3\,h^{-1}{\rm Mpc}$. Training such a model is possible because we do know the cosmology of each graph, but it is also possible to use it with real survey data by assuming the underlying cosmology of our Universe is the one measured by Planck. The performance of the model is shown in the bottom right panel of Figure~\ref{fig:inference_LH}. The accuracy of the model is significantly improved by the use of this global attribute: the value of the coefficient of determination is $R^2=0.75$, being much closer to 1 than in all the other cases studied with the LH. The precision is also significantly improved and equal to $\pm 0.029$. The precision obtained from the BAO+$f_i$ fit in~\cite{Setareh_2022} yields $RMSE=0.39$, once it has been re-scaled to the volume of the sub-boxes considered for this study (and all those implementing the \emph{fixed-$N_{\rm halo}$} method). This means that the interloper fraction predicted by this GNN model is 13 times more precise than that of the BAO+$f_i$ fit, while allowing for both cosmology and halo bias to vary. 

We consider the case where the posterior distributions of the cosmological parameters from which we draw the guessed cosmology are broader. We consider them to be a Gaussian distribution with a standard deviation equal to $2\times$ and $3\times$ the $1\sigma$ error reported from Planck (the precedent discussed results used $1\times$ the $1\sigma$ error). We trained such models where graphs have these updated global attributes and chose the model with the best validation loss. As expected, having a less precise determination of the cosmology of a graph worsens the performance of the GNN model: the $RMSE$ increases by $3\%$ and $14\%$ when considering $2\times$ and $3\times$ broader posteriors for the cosmological parameters, respectively, while the value of $R^2$ decreases by $3\%$ and $9\%$ in the two cases. Even if the posterior is broader by $3\times$, the GNN model performs better than when considering the monopole of the power spectrum as global attribute, and much better than without any global attribute.

\section{Testing the framework of the GNN approach}
\label{sec:discussion}

\subsection{Large scale information}

\begin{figure}[!t]
\centering
\includegraphics[width=.4\textwidth]{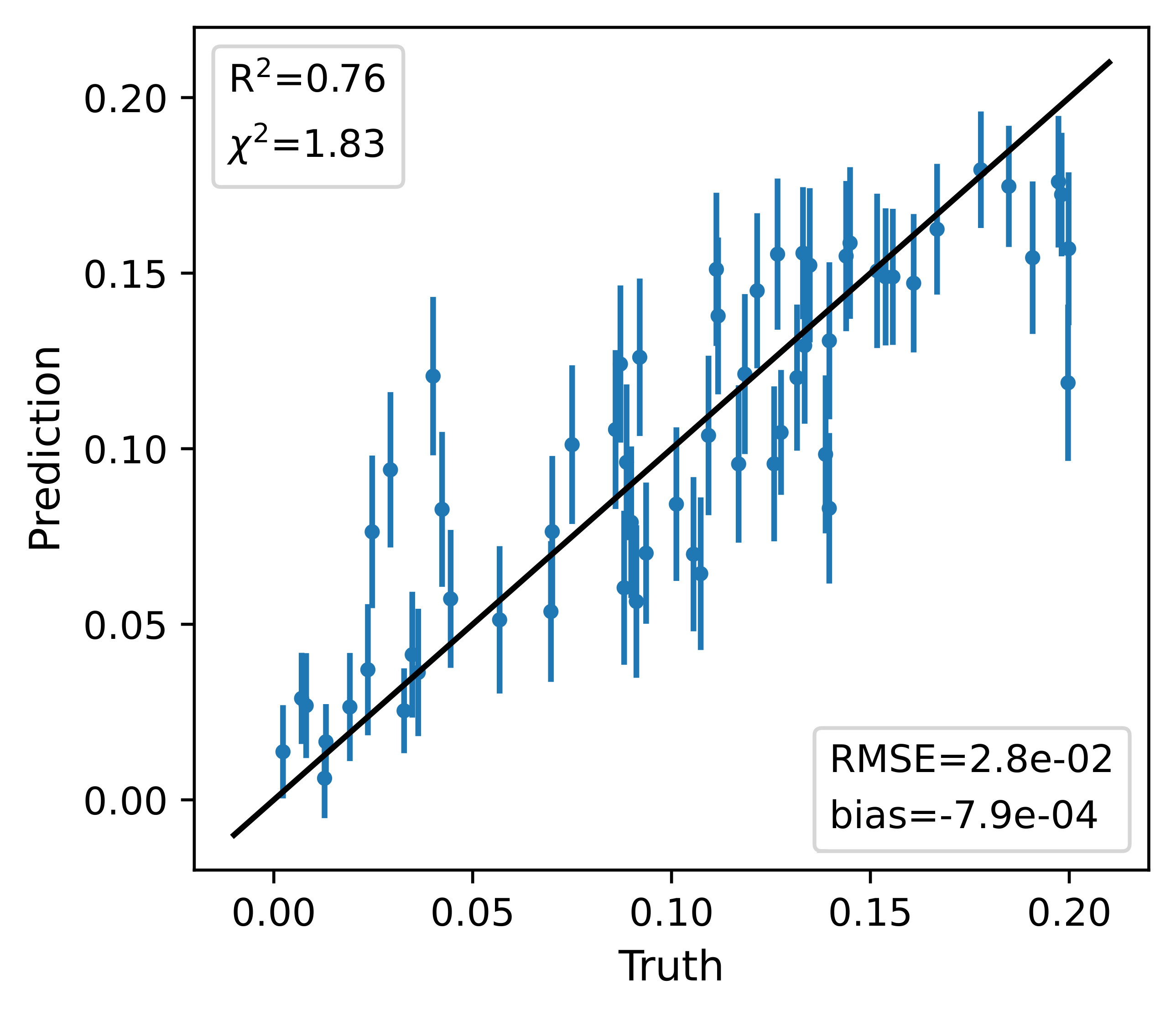}
\caption{Likelihood-free inference of the fraction of interlopers in redshift space halo catalogs built from the full boxes of the Latin Hypercube ({\tt SET2}). The halo bias scheme has been implemented using the \emph{fixed-$N_{\rm halo}$} method.
\label{fig:inference_LH_fullbox}}
\end{figure}

When training a GNN model on the Latin Hypercube ({\tt SET2}), we observed that the information coming from small scales does not seem to be sufficient to accurately and precisely predict the mean and standard deviation of the posterior distribution of the interloper fraction. In order to understand if including the information from large scales can ameliorate the GNN model, we generate a new data set where a graph is built using the full simulation box of size $1\,h^{-3}{\rm Gpc}^3$, and the halo bias scheme is implemented using the \emph{fixed-$N_{\rm halo}$} method with $N_{\rm halo}=4,500$. The number of halos is the same as the one previously used in $1/16$ of the full simulation volume, thus here the sample is less dense, allowing us to explore larger linking radii without running out of GPU memory. We perform a study using this setup applied to the LH and find that all the models with best validation loss have large linking radius, $r_{\rm link}\sim 100\,h^{-1}$Mpc. The best model has hyperparametrs $N_{\rm block}=1$, $N_{\rm hid}=21$, $l_r=1.7\times 10^{-4}$, $w_d=2.8\times 10^{-7}$, $r_{\rm  link} =100.2\,h^{-1}{\rm Mpc}$. Its performance on the test set is shown in Figure~\ref{fig:inference_LH_fullbox}, and the performance metrics are comparable to those obtained with graphs with global attribute being a guess for the cosmological parameters drawn from a Gaussian posterior with standard deviation equal to the $1\sigma$ reported by the Planck collaboration. This suggests that allowing the inclusion of large scales information is desirable when the training set exhibits variation of cosmology and halo bias scheme. Exploring large values of $r_{\rm link}$ while maintaining the full halo catalog has not been possible in this work because of the limited GPU memory. A promising avenue to include large-scale information while assuring for a small enough number of edges in each graph is represented, for example, by hierarchical GNNs~\cite{Sobolevsk_2021}.

\subsection{Spatial information}

\begin{figure}[t]
\centering
\includegraphics[width=.99\textwidth]{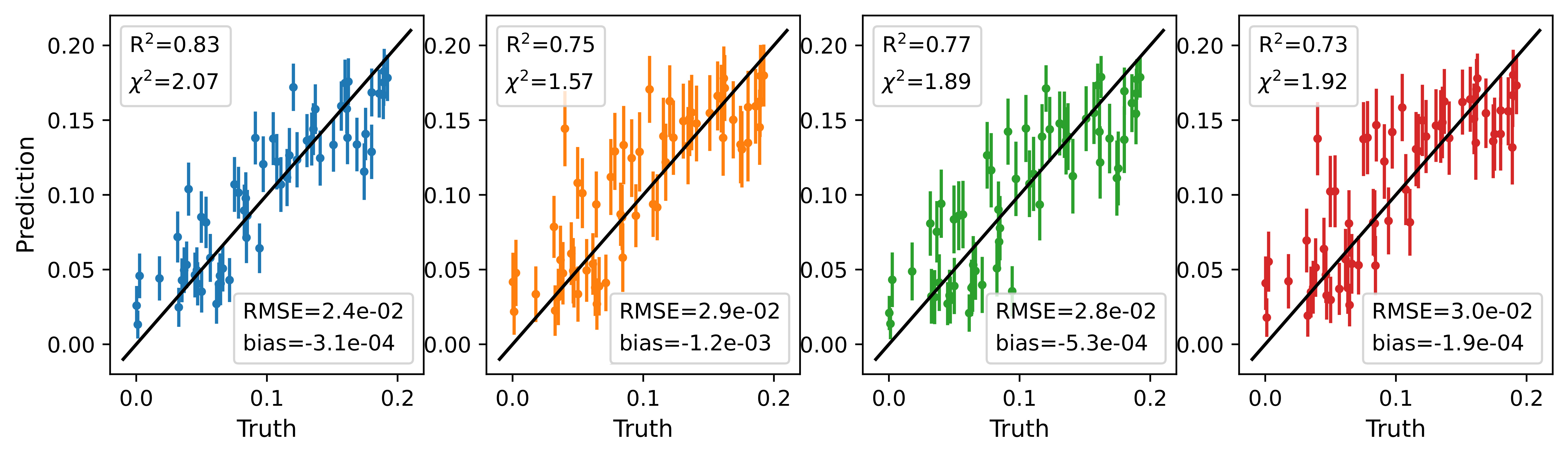}
\caption{Likelihood-free inference of interloper fractions in the test sets built using the fiducial cosmology simulations ({\tt SET1}) with \emph{fixed-$N_{\rm halo}$} bias scheme. From left to right, each plot shows the cases where different edge attributes have been implemented in the graphs: ${\bf e}^1_{ij} = \left[ r_\parallel , r_\perp \right]$ (blue), ${\bf e}^2_{ij} = \left[ r_\parallel \right]$ (orange), ${\bf e}^3_{ij} = \left[ r_\perp \right]$ (green), ${\bf e}^4_{ij} = \left[ \cos \theta \right]$ (red). 
\label{fig:inference_spacial_info}}
\end{figure}

In this work, we built graphs so that the edge attribute consists of 3 scalars that respect the symmetry of the problem. We want to understand if one or two of them carry most of the information about the amount of interlopers in a catalog, allowing us to obtain a GNN model with similar performance but operating on graphs with smaller (from the memory point of view) edge attributes. To investigate this, we consider four different cases where the edge attributes of each graph are 
\begin{equation}
    {\bf e}^1_{ij} = \left[ r_\parallel , r_\perp \right]\,,
    \qquad
    {\bf e}^2_{ij} = \left[ r_\parallel \right]\,,
    \qquad
    {\bf e}^3_{ij} = \left[ r_\perp \right]\,,
    \qquad
    {\bf e}^4_{ij} = \left[ \cos \theta \right]\,.
\end{equation}
We construct these four different sets of graphs using sub-boxes of size $250\times 250 \times 1,000 \, ({\rm Mpc}/h)^3$ along the $\hat{x}$, $\hat{y}$, and $\hat{z}$ directions and obtained from the 100 simulations in {\tt SET1}, with \emph{fixed-$N_{\rm halo}$} halo bias scheme that allows us to have the same halo number density in all sub-boxes. Figure~\ref{fig:inference_spacial_info} shows the performance of the best models on the test sets. The left panel displays the results for ${\bf e}^1$, where both distances along and perpendicular to the line-of-sight are encoded in the initial edge attribute. In this case, the best model exhibits values for the coefficient of determination, $R^2=0.83$, and the root mean square error, $RMSE=0.029$, that are very similar to those reported in the right panel of Figure~\ref{fig:inference_fid_varyb}, where the same setup is considered except for the inclusion of $\cos \theta$ in the edge attribute. This indicates that most of the information about the interloper fraction can be extracted using only distances between connected nodes. 

The second, third, and fourth panels in Figure~\ref{fig:inference_spacial_info} show the results when using ${\bf e}^2$, ${\bf e}^3$, and ${\bf e}^4$, respectively. In these cases, the performances of the best models are degraded, as an increase in the $RMSE$ values and a decrease in the coefficients of determination indicate. Moreover, while the best models for ${\bf e}^2$ and ${\bf e}^4$ have $r_{\rm  link}\sim 20-30\,h^{-1}{\rm Mpc}$ and use 1 or 2 GNN layers, the one for ${\bf e}^3$ has a very small linking radius, $r_{\rm  link}\sim 7\,h^{-1}{\rm Mpc}$, and uses only 1 GNN layer, indicating that is it taking information from extremely small scales only. 

From this analysis, two key messages emerge: eliminating $\cos \theta$ from the edge attributes might be a good choice to make the graphs less heavy from the memory point of view, and using only one of the three scalars as edge attribute degrades the GNN performance.

\section{Conclusions}
\label{sec:conclusions}
GNNs are designed to handle sparse and irregular data and are therefore a compelling approach for various analyses of galaxy or halo catalogs. In this paper, we showed that they can solve problems where one needs to identify a sub-sample of objects whose spatial properties differ from that of the core sample, such as interloper galaxies. We worked with cosmological simulations using halo positions to mimic the positions of galaxies in a real survey. In each halo catalog, we introduced \oiii-H$\beta$-like interlopers by displacing randomly selected halos along the line-of-sight, and we created different levels of contamination by randomly choosing the fraction of interloper within the range $f_i\in [0.0-0.1]$. Using the contaminated catalogs, we built graphs to describe their 3D spatial distribution: halos are nodes that are connected via edges only if their distance is smaller or equal to the linking radius $r_{\rm link}$ that is a hyperparameter. The spatial information is encoded in the edge attribute while respecting the symmetries of the problem. 

We considered the easiest case where GNNs perform likelihood-free inference of the interloper fraction in catalogs sharing the same underlying cosmology and the same halo bias scheme. In this case, using a small volume equal to $0.0225 h^{-3}{\rm Gpc}^3$, a GNN can accurately predict the mean of the posterior distribution of interloper fraction without any bias and with a precision equal to $0.015$, corresponding to the $15\%$ of the mean value considered: $f_i=0.1$. The analysis also showed that the GNN uses small-scale information, which is typically not used for cosmological constraints, to determine the interloper fraction.

Subsequently, we studied a more complicated task where the training set was built using a shared cosmology but different halo bias schemes. We implemented this scenario in two different ways: the \emph{varied-}$N_{\rm halo}$ and the \emph{fixed-}$N_{\rm halo}$ methods, where the number of halos is varied or fixed among the different catalogs. In both cases, the GNN gave unbiased results using small-scale information only, but the accuracy and precision were worse than in the scenario where all catalogs shared also the same halo bias. 

Finally, we considered the task that is more similar to the application to real data, but it is also the most difficult: predicting the interloper fraction while marginalizing over cosmology and halo bias, i.e. the training is performed using halo catalogs with varying cosmology and halo bias. In this case, a GNN that uses only small-scale information has limited capacity in estimating the mean and standard deviation of the posterior distribution of the interloper fraction. This might be due to degeneracies between cosmology and halo bias appearing on small scales. We tested this hypothesis by introducing large-scale information in different ways. First, we considered adding the monopole of the power spectrum measured from each catalog as a global attribute of each graph. In this case, the gain in the performance of the GNN was very marginal, probably due to the large cosmic variance in the monopole. Second, we considered adding a guess for the 5 cosmological parameters as the global attribute of each graph. In this case, the performance of the GNN improved considerably, reaching similar accuracy and precision as the scenario where only the halo bias was varied, but the cosmology was kept fixed among the catalogs. Lastly, we considered adding large-scale information by allowing for a large linking radius, $r_{\rm link}<120\,h^{-1}{\rm Mpc}$. In order to deal with such large linking radii while avoiding heavy graphs with many edges, we considered larger volumes but sub-sampled the number of halos. In this case, the performance of the GNN is considerably ameliorated compared to the case where only small-scale information was used, giving a precision and an accuracy that are better than in the scenario where small linking radii and a global attribute with a guess for the cosmology were used. 

We also investigated the information contained in the three scalars that make up the edge attributes. We found that each singular scalar is not sufficient to learn the mean and standard deviation of the posterior distribution of interloper fraction as precisely as when using all of them. However, $r_\parallel$ and $r_\perp$---the size of the edge perpendicular and parallel to the line-of-sight---seem to encode most of the information. Therefore, in an effort to make the graphs less heavy, it might be a good choice to use only these two scalars as edge attributes. 

Overall, GNNs have proven to be valuable methods for inferring the interloper fraction in a catalog. They outperformed more standard methods, such as BAO+$f_i$ fitting functions, in measuring the interloper fraction with higher precision. On the other hand, when applied to data from galaxy surveys, additional work needs to be done when building the training set, which includes modeling the Finger-of-God effect, the survey geometry, and the systematics of the survey. 

It is worth noticing that, although we have worked in comoving coordinates and assumed a fiducial cosmology to perform the analysis, GNNs can in principle work in observational space, i.e. the GNNs can be trained directly in (RA, DEC, z) rather than comoving coordinates. This can be a major advantage, since it does not require assuming a fiducial cosmology; on the other hand, its implementation is not straightforward since it requires a different procedure on building the graphs and in particular on deciding if two nodes are connected via an edge. Moreover, GNNs have shown to be powerful tools to constrain cosmology using information beyond the two-point function~\cite{Villanueva-Domingo_2022b,DeSanti_2023}. A further generalization of the method proposed in this paper could consist of the simultaneous inference of interloper fraction and cosmological parameters.
\appendix
\section{Epistemic Error}
\label{sec:epistemic}
\begin{figure}[!t]
\centering
\includegraphics[width=.99\textwidth]{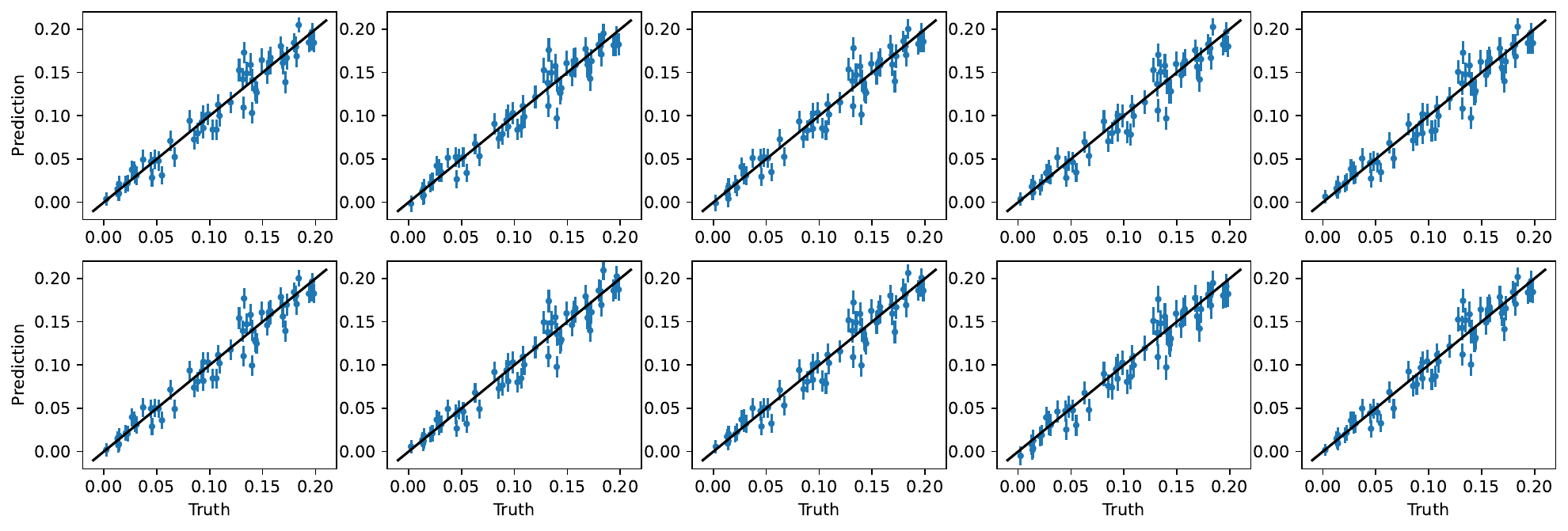}
\caption{Likelihood-free inference of interloper fractions in the test sets using the model in the right panel of Figure~\ref{fig:inference_fid_fixb}. Each panel shows the model trained using different random seeds to initialize the weights of the GNNs.   
\label{fig:epistemic}}
\end{figure}

We evaluate the epistemic error---the error of the GNN model itself---by retraining 10 times a model with given hyperparameters; each time, the initial weights of the MLPs are drawn using a different random seed. We perform this test in the case where both cosmology and halo bias scheme are fixed across the training set, i.e. for the case discussed in Section~\ref{sec:inference_fid_fixb} and in particular for the redshift-space scenario. Figure~\ref{fig:epistemic} displays the inference predicted by these 10 different retrained models on the test set. We can see that the results are very similar, almost indistinguishable, among different retrained models. We quantify the epistemic error as,
\begin{equation}
    \epsilon_e = \frac{1}{N_\mathcal{G}}\sum_{g \in \mathcal{G}} \left[\frac{1}{10} \sum_{j=1}^{10} (\mu_{g,j} - \bar{\mu}_g)^2\right]^{1/2}
\end{equation}
where, first, we compute the standard deviation (among different trained models $j$) of the predicted interloper fraction $\mu$ for a given graph $g$ in the test set $\mathcal{G}$, then we average the standard deviation across different graphs. The resulting epistemic error is $\epsilon=0.002$, which is about $15\%$ of the standard deviation of the posterior distribution of interloper fractions.

% Bibliography

%% [A] Recommended: using JHEP.bst file
\bibliographystyle{JHEP}
\bibliography{biblio.bib}
\end{document}